# Spin frustration and unconventional spin twisting state in van der Waals ferromagnet/antiferromagnet heterostructures

Tianye Wang[1,2], Qian Li[3], Mengmeng Yang[4], Yu Sun[5], Alpha T. N'Diaye[2], Christoph Klewe[2], Andreas Scholl[2], Xianzhe Chen[2,6], Xiaoxi Huang[6], Hongrui Zhang[2,6,7], Santai Yang[1], Xixiang Zhang[8], Chanyong Hwang[9], Padraic C. Shafer[2], Michael F. Crommie[1,2,10], Ramamoorthy Ramesh[1,2,6], Zi Q. Qiu[1,2,*]

1. Department of Physics, University of California, Berkeley, CA, USA
2. Materials Sciences Division, Lawrence Berkeley National Laboratory, Berkeley, CA, USA.
3. National Synchrotron Radiation Laboratory, and School of Nuclear Science and Technology, University of Science and Technology of China, Hefei, Anhui, China
4. Anhui Key Laboratory of Magnetic Functional Materials and Devices, Institute of Physical Science and Information Technology, Anhui University, Hefei, China
5. School of Physical Science and Technology, ShanghaiTech University, Shanghai, China
6. Department of Materials Science and Engineering, University of California, Berkeley, CA, USA.
7. Ningbo Institute of Materials Technology & Engineering, Chinese Academy of Sciences, Ningbo, China
8. Physical Science and Engineering Division (PSE), King Abdullah University of Science and Technology (KAUST), Thuwal, Saudi Arabia
9. Korea Research Institute of Standards and Science, Yuseong, Daejeon, Korea
10. Kavli Energy Nano Sciences Institute, University of California Berkeley and Lawrence Berkeley National Laboratory, Berkeley, CA, USA.

* Corresponding author. Email: qiu@berkeley.edu

## Abstract

Atomically flat surfaces of van der Waals (vdW) materials pave an avenue for addressing a long-standing fundamental issue of how a perfectly compensated antiferromagnet (AFM) surface frustrates a ferromagnetic (FM) overlayer in FM/AFM heterostructures. By revealing the AFM and FM spin structures separately in vdW $Fe_5GeTe_2$/$NiPS_3$ heterostructures, we find that C-type in-plane AFM $NiPS_3$ develops three equivalent AFM domains which are robust against external magnetic field and magnetic coupling with $Fe_5GeTe_2$. Consequently, spin frustration at the $Fe_5GeTe_2$/$NiPS_3$ interface was shown to develop a perpendicular $Fe_5GeTe_2$ magnetization in the interfacial region that switches separately from the bulk of the $Fe_5GeTe_2$ magnetizations. In particular, we discover an unconventional spin twisting state that the $Fe_5GeTe_2$ spins twist from perpendicular direction near the interface to in-plane direction away from the interface in $Fe_5GeTe_2$/$NiPS_3$. Our finding of the twisting spin texture is a unique property of spin frustration in van der Waals magnetic heterostructures.

# Introduction

The discovery of magnetic order in van der Waals (vdW) materials has opened up unprecedented opportunities for spintronics research [1] [2] [3]. Such two-dimensional layered structures and atomically flat surfaces of vdW materials make it possible to address some fundamental issues which could not be resolved in conventional magnetic materials. One such long-standing issue known as the so-called spin frustration, which arises at the interface between a fully compensated antiferromagnet(AFM) surface (e.g., C- or G-type AFM) and a ferromagnet(FM), is critically important in both technological applications [4] [5] [6] [7] [8] and fundamental understanding of magnetic systems [9] [10] [11] [12]. The zero net magnetic moment at the AFM surface prohibits the minimization of the exchange interactions for all pairs of spins at the FM/AFM interface. In FM/AFM heterostructures of conventional magnetic materials, the inevitable imperfection of the FM/AFM interface (roughness, defects, and interdiffusion, etc.) usually creates uncompensated spins at the AFM surface to overwhelm the intrinsic spin frustration, although such extrinsic uncompensated spins are critical to the exchange bias effect[13]. The atomically flat vdW magnetic materials provide an ideal surface offering a unique opportunity for the study of spin frustration. Indeed, recent studies on vdW FM/AFM systems [14] [15] [16] [17] [18] [19] already revealed some exciting phenomena that are very different from conventional FM/AFM systems. Despite the early interests, these studies also face obstacles such as oxidized AFM layers which consist of imperfect interfaces [14] [15] [20] [21] [22] or A-type AFM materials whose surface spins are intrinsically uncompensated in terms of the FM/AFM interfacial interaction [16] [19] [23] [24] [25] [26]. Moreover, it is challenging to directly detect the AFM order in vdW FM/AFM systems, making it difficult to relate the FM behaviors with the AFM spin structures, and hence, vdW AFM with strong perpendicular magnetic anisotropy (PMA) are usually employed to avoid ambiguity of the AFM orders [18] [27] [28] [29] [23] [24] [25]. All the above limitations can be overcome by studying FM / (in-plane C-type AFM) vdW heterostructures which contain the key component of spin-flop transition in spin frustration.

In this work, we investigate $Fe_5GeTe_2$/$NiPS_3$ vdW FM/AFM heterostructures by measuring both AFM and FM spin structures. $Fe_5GeTe_2$ is a vdW FM whose magnetization undergoes a spin-reorientation transition from out-of-plane magnetization at smaller thicknesses to in-plane magnetization at larger thicknesses [30] [31]. $NiPS_3$ is an in-plane C-type vdW AFM with a uniaxial magnetic anisotropy along the a-axis of its monoclinic lattice [32] [33]. Therefore, atomically flat vdW $Fe_5GeTe_2$/$NiPS_3$ heterostructures provides an ideal platform for the investigation of spin frustration in fully compensated in-plane AFM surface coupled with both out-of-plane and in-plane magnetized FM layers.

## Results

**Direct element-specific characterization of the AFM order in NiPS$_3$**

To realize element-specific measurement of the AFM order in heterostructures, we first characterized macroscopic X-ray Magnetic Linear Dichroism (XMLD) on a bulk NiPS$_3$ crystal. The measurement geometry is depicted in Fig. 1a. Fig. 1b shows the NiPS$_3$ X-ray Absorption Spectra (XAS) and XMLD at T = 23K. The temperature-dependent XMLD (Fig. 1c) exhibits a clear non-zero value at low temperatures and drops to zero near the NiPS$_3$ Néel temperature $T_N = $ 155K [32], showing the AFM origin of the XMLD signal and the in-plane AFM spin axis in our bulk NiPS$_3$ sample.

Using XMLD- Photoemission electron microscopy (XMLD-PEEM), we directly imaged the AFM domains in a ~70nm-thick tape-exfoliated NiPS$_3$ flake. Fig. 1d exhibits multiple AFM domains at T=106K with three different magnetic contrasts corresponding to the three equivalent AFM ordering axes of a vdW hexagonal monolayer lattice.

We further characterized the NiPS$_3$ XMLD in a micron-sized Fe$_5$GeTe$_2$(35nm)/NiPS$_3$(31nm) heterostructure using the membrane window technique introduced in a later section as well as in the Methods. Results and analysis are shown in Fig. S1. We found that the local magnetic anisotropies of NiPS$_3$ are strong enough against spin-flop transition under the 0.4T cooling field plus the Fe$_5$GeTe$_2$/NiPS$_3$ interfacial magnetic coupling. This behavior is very different from conventional FM/AFM systems where the interfacial coupling could align the AFM Néel vector after field cooling [34] [35].

**Out-of-plane Fe$_5$GeTe$_2$(16nm) magnetization with and without NiPS$_3$**

Out-of-plane magnetized vdW materials [36] [37] [38] [39] and the manipulation of their magnetic properties (e.g., thin Fe$_5$GeTe$_2$ with PMA) have been reported extensively [40] [41] [42]. Transport measurement via anomalous Hall effect (AHE) has been widely used for this type of study to provide fast and direct representation of out-of-plane hysteresis loops. As reported in the literature [30], Fe$_5$GeTe$_2$ consists of a spin reorientation transition (SRT) from out-of-plane magnetization at smaller thicknesses to in-plane magnetization at larger thicknesses. We first present our transport study using AHE on a thin 16nm-thick $Fe_5GeTe_2$, half of which was covered by $NiPS_3$ flake as shown in Fig. 2a. The comparison between the $Fe_5GeTe_2$ region and the $Fe_5GeTe_2/NiPS_3$ heterostructure region using one same $Fe_5GeTe_2$ flake rules out the sample-to-sample variation which could show up otherwise from multiple samples.

The sample was cooled down to 80K within a 9T out-of-plane magnetic field to ensure a single domain of the $Fe_5GeTe_2$ perpendicular magnetization. The square-shaped out-of-plane loops (Fig. 2b-c) from both the $Fe_5GeTe_2/NiPS_3$ heterostructure and the $Fe_5GeTe_2$ single flake show that the 16nm-thick $Fe_5GeTe_2$ has a perpendicular magnetization. Unlike the single flake $Fe_5GeTe_2$ loop which exhibits a sharp one-step magnetization switching at the coercivity, the $Fe_5GeTe_2/NiPS_3$ heterostructure loop exhibits roughly a two-step switching with one switching identical to single layer $Fe_5GeTe_2$ and the other at a higher coercive field, showing that the higher field switching must originate from the $Fe_5GeTe_2/NiPS_3$ interfacial spin frustration. Considering the fact that vdW magnet has a much weaker interlayer interaction than intralayer interaction, we attribute the two-step switching in the $Fe_5GeTe_2/NiPS_3$ loop to the mechanism that the $Fe_5GeTe_2/NiPS_3$ interfacial spin frustration induces an additional interfacial PMA which enhances the out-of-plane switching field of the $Fe_5GeTe_2$ in the interfacial region while leaving the rest $Fe_5GeTe_2$ unaffected in the region away from the interface. Quantitative analysis will be presented in Discussion.

**In-plane $Fe_5GeTe_2$ magnetization in $Fe_5GeTe_2$(30~35nm)/$NiPS_3$**

Since AHE measurement is not the best technique for in-plane magnetizations, we fabricated $Fe_5GeTe_2$(30nm) single flake and $Fe_5GeTe_2$(35nm)/$NiPS_3$(31nm) heterostructure samples and measured their hysteresis loops using element-resolved XMCD measurement with the membrane window technique shown in Fig. 3a-b (also see Methods). XMCD measures the projection of magnetization along the X-ray direction, we characterized the in-plane magnetic hysteresis loops using 25° X-ray incident angle (Fig. 3c-d) which will also involve the out-of-plane component if there is any. After a 1.8T in-plane field cooling, the $Fe_5GeTe_2$(35nm)/$NiPS_3$(31nm) heterostructure exhibits the same in-plane easy-axis loop as the 30nm-$Fe_5GeTe_2$ and 51nm-$Fe_5GeTe_2$ single flake (Fig. S2b) at temperatures higher than the $NiPS_3$ Néel temperature, consistent with the published result that thicker $Fe_5GeTe_2$ film has an in-plane magnetization [30]. Below the $NiPS_3$ Néel temperature, the $Fe_5GeTe_2$(35nm)/$NiPS_3$(31nm) heterostructure gradually exhibits a two-step switching hysteresis loop, indicating the effect of $NiPS_3$ AFM orders. Here we discuss two important observations. First, no exchange bias was observed after the in-plane magnetic field cooling, unlike conventional FM/AFM systems. The exchange bias in a FM/AFM system should be absent for perfectly compensated AFM surface. We want to point out that the existence of exchange bias in experimental FM/AFM systems is due to the inevitable uncompensated AFM spins at or near the interface (e.g., caused by defect, interfacial roughness, and chemical imbalance, etc.) [13] [43], and hence, the absence of a measurable exchange bias in our sample is consistent with a fully compensated AFM order of the atomically flat vdW $NiPS_3$ at the $Fe_5GeTe_2$/$NiPS_3$ interface as expected from the vdW nature of $NiPS_3$. Second, the $Fe_5GeTe_2$/$NiPS_3$ hysteresis loop below the $NiPS_3$ Néel temperature (Fig. 3c) is roughly a superposition of the $Fe_5GeTe_2$ single flake loop and the other loop with a higher switching field, similar to the perpendicular magnetized 16nm-thick $Fe_5GeTe_2$/$NiPS_3$ in Fig. 2b with the extra switching step beyond the coercivity. Based on our earlier analysis of Fig. 2b that the interfacial region and the region away from the interface switch separately, we believe that the $Fe_5GeTe_2$ in the interfacial region in $Fe_5GeTe_2$(35nm)/$NiPS_3$ should also have a perpendicular magnetization due to the $Fe_5GeTe_2$/$NiPS_3$ interfacial spin frustration even though the region away from the interface has an in-plane magnetization (More comprehensive analysis will be shown later). In other words, the spin frustration at the $Fe_5GeTe_2$(35nm)/$NiPS_3$(31nm) interface result in a twisting spin texture of the $Fe_5GeTe_2$ from perpendicular magnetization in the interfacial region to an in-plane magnetization in the region away from the interface in the 35nm $Fe_5GeTe_2$.

To verify the interfacial perpendicular $Fe_5GeTe_2$ magnetization in (thick-$Fe_5GeTe_2$)/$NiPS_3$ heterostructures, we fabricated another $Fe_5GeTe_2$(31nm)/$NiPS_3$ sample due to limited beamtime for the previous sample, and measured XMCD hysteresis loops (Fig. 3e-g). We first confirmed that the 25° hysteresis loop at 80K (Fig. 3e) exhibits the same two-step switching process as the previous sample in Fig. 3c. Then we performed XMCD measurement at 0° X-ray incident angle (Fig. 3f) which only picks up the out-of-plane magnetic component. The 0° loop exhibits an open loop with only one-step switching at ~0.1T field which corresponds to the second-step switching of the 25° loop in Fig. 3e. By subtracting the out-of-plane loop from the 25° loop with the factor cos(25°), we constructed the projection of the pure in-plane loop (Fig. 3g) which shows the same in-plane easy-axis loop of single-flake $Fe_5GeTe_2$ as in Fig. 3c-d and Fig. S2b. This result further confirms that the second-step switching in the 25° loops originates from an out-of-plane magnetization which coexists with the in-plane magnetizations in the $Fe_5GeTe_2$. To have more comprehensive evidence, we further characterized the angular dependence of the $Fe_5GeTe_2$ magnetization by measuring the XMCD magnitude as a function of the X-ray incident angle (Fig. 3h-i). The sample was cooled from room temperature within a -1.8 T field along $\theta = -45°$ off-normal direction to wipe out magnetic domains. The field was then turned off (except a tiny remanent field of the electromagnet) so that the Fe$_5$GeTe$_2$ out-of-plane magnetization in the interfacial region and the in-plane magnetization of the region away from the interface should go back to their own easy-axis directions, forming a total magnetization M with an angle $\alpha = tan^{-1}(d_\perp/d_\parallel)$ tilting out of the film plane (the small remanent field prevents the degenerate state of $M_\perp < 0$), where $d_\perp$ and $d_\parallel$ are the thicknesses of the interfacial region and the region away from the interface, respectively. As the X-ray incident angle $\theta$ changes, the XMCD magnitude which is proportional to the projection of the total magnetization along the X-ray direction should depend on $\theta$ with $|XMCD|\sim|sin(\theta - \alpha)|$. Then for a non-zero out-of-plane magnetization ($\alpha \neq 0$), the XMCD magnitude should reach zero at $\theta = \alpha$ as opposed to $\theta = 0°$ for a perfectly in-plane magnetization ($\alpha = 0$). At T=190K, the XMCD magnitude reaches zero at $\theta \approx 0°$ (Fig. 3i), showing a perfectly in-plane Fe$_5$GeTe$_2$ magnetization above the NiPS$_3$ Néel temperature. At T=80K, the XMCD magnitude reaches zero at $\theta \approx 12°\sim14°$, showing the existence of an out-of-plane Fe$_5$GeTe$_2$ magnetic component below the NiPS$_3$ Néel temperature. Quantitative analysis will be presented in Discussion.

**Fe$_5$GeTe$_2$(22nm and 31nm in one flake)/NiPS$_3$ heterostructure**

We further studied the thickness-dependent hysteresis loops (Fig. 4) by fabricating a Fe$_5$GeTe$_2$/NiPS$_3$ heterostructure in which the Fe$_5$GeTe$_2$ flake contains two different thickness regions (22nm and 31nm). This heterostructure guarantees an identical Fe$_5$GeTe$_2$/NiPS$_3$ interface to eliminate sample-dependent variations. Recalling that the X-ray spot size is greater than the membrane window size and that the transmission X-ray measures the signals from the 30μm whole window region, the temperature-dependent hysteresis loops of the whole Fe$_5$GeTe$_2$/NiPS$_3$ (including both 22nm and 31nm Fe$_5$GeTe$_2$ thicknesses) were measured at 0° and 25° X-ray incident angles after a 1.8T in-plane field cooling from room temperature down to 80K. (Comparison between in-plane and out-of-plane field cooling is shown in Fig. S3.)

Above the NiPS$_3$ Néel temperature, the Fe$_5$GeTe$_2$(22nm, 31nm)/NiPS$_3$ 25° loop (Fig. 4b, 180K) consists of two switching fields with the smaller one corresponding to the in-plane magnetization from the Fe$_5$GeTe$_2$(31nm)/NiPS$_3$, and the higher one corresponding to the out-of-plane magnetization from the Fe$_5$GeTe$_2$(22nm)/NiPS$_3$, evidenced by 0° loop in Fig. 4a. The constructed projection of in-plane hysteresis loops (Fig. 4c, same way as Fig. 3g) indeed exhibits only one switching field as an in-plane magnetized Fe$_5$GeTe$_2$ flake. Below the NiPS$_3$ Néel temperature, the projection of in-plane hysteresis loop (Fig. 4c) still exhibits the shape of in-plane magnetized single

Fe$_5$GeTe$_2$ flake without much increase in height. The 0° loop, however, shows an obvious increase of the out-of-plane magnitude compared to 180K, indicating that the interfacial region of the 31nm-thick Fe$_5$GeTe$_2$ switched its magnetization from in-plane direction at 180K to out-of-plane direction at 80K. Most importantly as shown in Fig. 4d, the temperature-dependent out-of-plane loop magnitude increases in the same way in which NiPS$_3$ XMLD signal increases as temperature decreases (red open circles). Such consistency proves that the increase in out-of-plane magnetic moments from the interfacial region of the 31nm Fe$_5$GeTe$_2$ originates from the FM/AFM interfacial spin frustration. A more quantitative analysis will be presented next.

## Discussion

**Anisotropic domain wall width in vdW ferromagnets and unconventional spin twisting state**

Ferromagnets can change spin orientation over the length scale of the domain wall width $L_{DW} \approx \sqrt{A/K}$, where $A$ is the magnetic exchange interaction along the width direction, and $K$ is the uniaxial magnetic anisotropy. In conventional magnetic materials, the domain wall width is on the order of $L_{DW} \sim 20\text{-}100$nm [44] so that spin twisting usually does not occur along the surface normal direction inside an ultrathin film. In vdW magnetic materials, however, the much weaker interlayer interaction could lead to ultrathin domain walls along the surface normal direction, making it possible to switch different regions along the surface normal direction of a FM vdW thin film separately. Fe$_3$GeTe$_2$ was reported to have only ~6nm in-plane domain wall width [45], implying that domain wall width along the surface normal direction could be even smaller than 6nm. Therefore, within the Fe$_5$GeTe$_2$ in the heterostructures, the interfacial region has a perpendicular magnetization due to the Fe$_5$GeTe$_2$/NiPS$_3$ spin frustration (see next subsection), while the region away from the interface has "intrinsic" magnetic properties of a single Fe$_5$GeTe$_2$ flake. These two regions correspond to the two-step switching processes observed in the hysteresis loops below NiPS$_3$ Néel temperature.

**Quantitative analysis of data based on the spin twisting state hypothesis**

In this section we present our quantitative analysis of all the samples. The results are highly consistent with each other, providing strong evidence for the conclusion of the spin twisting state. We start the quantitative analysis from the hysteresis loop of Fe$_5$GeTe$_2$(16nm) in Fig. 2b where both the interfacial region and the region away from the interface have perpendicular magnetizations. Fig. 2b shows that the interfacial region accounts for ~35% of the total hysteresis loop magnitude, corresponding to ~5.5nm thickness. Then, considering the physical picture shown by Fig. 5, the 35nm-thick $Fe_5GeTe_2$ sample in Fig. 3c should also consist of ~5.5nm-thick interfacial region with perpendicular magnetization, leaving the rest ~29.5nm-thick region with an in-plane magnetization. Projecting the magnetizations of these two regions to the 25° X-ray direction, the ~5.5nm-thick out-of-plane magnetization should contribute $\frac{5.5cos25°}{5.5cos25°+29.5sin25°} \sim 29\%$ of the total XMCD signal, in agreement with the ~30% obtained from Fig. 3c. This analysis also applies to the $Fe_5GeTe_2$(31nm) hysteresis loops in Fig. 3e-g. Moreover, with separating the 31nm-thick $Fe_5GeTe_2$ film into 5.5nm-thick out-of-plane magnetization and 25.5nm-thick in-plane magnetization, the ratio of these two regions $\frac{5.5}{25.5} = 0.22$ is very close to $\tan(12°\sim14°) = 0.21\sim0.25$ from Fig. 3i. Finally, for the $Fe_5GeTe_2$(22nm and 31nm) in Fig. 4d, we extrapolate the temperature-dependent out-of-plane magnitude from high temperature to low temperature (green dashed line). The actual XMCD at 80K has a value of 0.063 which is 37% greater than the extrapolated value of 0.046. Assuming that this increment comes from magnetic moments in the interfacial region of the 31nm $Fe_5GeTe_2$, we estimate the thickness of this interfacial region to be $\frac{0.37 \times 22}{1.44} = 5.7$nm which is very close to the 5.5nm thickness obtained in previous analysis.

The consistency of the ~5.5nm-thick interfacial region in all samples strongly support our conclusion that spin frustration in $Fe_5GeTe_2$/$NiPS_3$ results in a perpendicular $Fe_5GeTe_2$ magnetization within ~5.5nm-thick interfacial region in both thick (in-plane) and thin (out-of-plane) $Fe_5GeTe_2$ samples. Particularly, thick $Fe_5GeTe_2$ film in $Fe_5GeTe_2$/$NiPS_3$ develops a new novel spin twisting state that the $Fe_5GeTe_2$ spins twist from out-of-plane direction in the interfacial region to in-plane direction away from the interface, an unconventional state that does not exist in conventional FM thin films.

**PMA induced by FM/AFM interfacial spin frustration**

It is known that interfacial electronic state in a FM layer on a substrate induces an interfacial magnetic anisotropy [46] [47] [48] which, in special cases, could lead to a PMA [49] [50] [51] [52] [53]. Although such interfacial magnetic anisotropy might also exist in our $Fe_5GeTe_2$/$NiPS_3$ system, the fact that the out-of-plane $Fe_5GeTe_2$ magnetization in the interfacial region is established below the NiPS$_3$ Néel temperature shows that it is the NiPS$_3$ AFM order that induces an additional PMA in the Fe$_5$GeTe$_2$, similar to the Co/Fe/e-fct-Mn system [54].

To better understand how the AFM surface results in the PMA, we propose our explanations illustrated as below. As presented in the Supplementary Materials, we describe two spin-flop-like effects of the magnetic interfacial spin frustration including a small angular twisting of the FM spins relative to the AFM spins and a small angular twisting of the AFM spins towards the FM magnetization. Both spin-flop-like effects tend to align the FM magnetization and the AFM Néel vector perpendicular to each other.

Then, it is the AFM property of NiPS$_3$ that further transforms such spin-flop-like effects into a PMA. In the case of a strong AFM anisotropy which fixes the AFM spin axis in space, the FM spin axis will rotate by 90º to the perpendicular direction to the AF spin axis [55]. Moreover, NiPS$_3$ is known to have relatively large magnetic anisotropy with a 6T spin-flop-transition field [33], which is why the NiPS$_3$ spins in Fe$_5$GeTe$_2$/NiPS$_3$ remain fixed along the three local in-plane easy axes after field cooling. The spin-flop-like effects in this case favor the interfacial Fe$_5$GeTe$_2$ magnetization to be out-of-plane in order to be perpendicular to all the AFM spins in the in-plane NiPS$_3$ domains. More importantly, it was shown that the magnetic susceptibility of NiPS$_3$ along out-of-plane direction is approximately 47% greater than those along the in-plane directions (literature [33] Fig. 2c, 2d(inset), and 2e). Such strong anisotropic AFM magnetic susceptibility can result in a PMA in adjacent Fe$_5$GeTe$_2$ even when the whole NiPS$_3$ sample is a single AFM domain, to lower the Zeeman energy during the spin-flop transition due to net magnetization of the AFM film.

## Summary

In summary, we found that spin frustration at the atomically flat interface of Fe$_5$GeTe$_2$/NiPS$_3$ heterostructures induces an interfacial perpendicular Fe$_5$GeTe$_2$ magnetization of ~5.5nm thickness. This interfacial out-of-plane magnetization not only switches separately from the region away from the interface, but also generates an unconventional magnetic state that the Fe$_5$GeTe$_2$ spins twist from out-of-plane direction near the interface to in-plane direction away from the interface in thicker-Fe$_5$GeTe$_2$/NiPS$_3$ samples. The twisting spin texture and the two-step separate magnetization switching are unique properties of vdW magnetic heterostructures which are absent in conventional magnetic materials.

# Methods

**XMLD of bulk NiPS$_3$ crystal**

XMLD measurements of bulk NiPS$_3$ were performed at the beamline 4.0.2 at Advanced Light Source (ALS) of Lawrence Berkeley National Lab (LBNL). A NiPS$_3$ bulk crystal was cleaved by a tape before being loaded into the vacuum chamber of the beamline. Linearly polarized X-ray was incident onto NiPS$_3$ crystal surface at normal incidence. XAS at Ni L$_2$-edges were obtained via the photoemission electron yield, and the XMLD was subsequently obtained by taking the difference of the XAS between two orthogonal X-ray linear polarizations.

**Measurements of micron-scale vdW samples using macroscopic XMCD and XMLD**

XMCD and XMLD experiments on micron-scale vdW samples were performed at beamline 6.3.1 and 4.0.2, respectively, at ALS of LBNL. To better single out the X-ray signals from micron-scale vdW flake samples especially from bilayer heterostructures, we transferred samples onto SiN$_x$ membrane window substrates using the PDMS-mediated dry transfer method [56]. The membrane window consists of 200μm-thick hollow Si frame with a 30μm × 30μm window at the center, and the whole substrate is covered by a 20nm-thick uniform SiN$_x$ membrane film on top including the hollow window. Because X-ray cannot transmit through the thick Si frame, the transmitted X-ray will be 100% through the flake sample on top of the membrane window. In this way, transmission X-ray intensity (normalized by incident X-ray intensity and the zero-dichroism L$_3$ absorption peak height) measures the flake samples with a greatly enhanced signal-to-noise ratio. This method is particularly important for stacked micron-sized bilayer heterostructures where the X-ray spot covers not only the bilayer region but also the single layer regions. **XMLD-PEEM imaging of NiPS$_3$ antiferromagnetic domains**

PEEM experiments were performed at the beamline 11.0.1 at ALS of LBNL. Linear polarized X-ray was incident onto NiPS$_3$ flakes at 60° incident angle, and the photoemission electrons were used to image the sample with electron microscopy. Division between images obtained with two mutually perpendicular X-ray linear polarizations reflects the antiferromagnetic domains of the sample. X-ray photon energy was fixed at 869.8eV which gives the maximum XMLD at the Ni L$_2$-edge.

**Local electronic transport measurement on $Fe_5GeTe_2$ partially covered by $NiPS_3$**

Six electrodes (one current line and two pairs of Hall bars) were fabricated on the $SiO_2$/Si wafer via photolithography. $Fe_5GeTe_2$/$NiPS_3$ heterostructure was transferred on top of the electrodes, in which the 16nm $Fe_5GeTe_2$ was partially covered by a 37nm-thick $NiPS_3$ flake via the PDMS dry transfer method. A thin h-BN was transferred on top as a protection layer. In this way, a comparison of the AHE measurements from the two halves of the sample could single out the $Fe_5GeTe_2$/$NiPS_3$ interfacial coupling effect by eliminating any artifacts due to sample-to-sample variations (e.g., differences in thickness, temperature and inhomogeneity, etc). The Hall resistance was measured at different temperatures by sweeping an out-of-plane magnetic field between $\pm\ 0.5$T.

**Sample thickness characterization**

Thicknesses of the vdW samples were characterized using tapping-mode atomic force microscopy. The membrane right over the 30μm × 30μm hollow Si window is soft and elastic, and the tapping tip would induce elastic deformation on the indirectly supported membrane during the measurement. Therefore, thicknesses were obtained from the flake edges outside the window region for precise results.

# Reference

[1] C. Gong, L. Li, Z. Li, et al., " Discovery of intrinsic ferromagnetism in two-dimensional van der Waals crystals.," *Nature,* vol. 546, pp. 265-269, 2017.

[2] K. S. Burch, D. Mandrus, and J.-G. Park, "Magnetism in two-dimensional van der Waals materials," *Nature,* vol. 563, pp. 47-52, 2018.

[3] H. Li, S. Ruan, and Y. -J. Zeng, "Intrinsic Van Der Waals Magnetic Materials from Bulk to the 2D Limit: New Frontiers of Spintronics.," *Adv. Mater.,* vol. 31, p. 1900065, 2019.

[4] J. Cramer, F. Fuhrmann, U. Ritzmann, et al., "Magnon detection using a ferroic collinear multilayer spin valve.," *Nat. Commun.,* vol. 9, p. 1089, 2018.

[5] B. Dieny, "Giant magnetoresistance in spin-valve multilayers.," *J. Magn. Magn. Mater.,* vol. 136, pp. 335-359, 1994.

[6]  X. Chen, T. Higo, K. Tanaka, et al., "Octupole-driven magnetoresistance in an antiferromagnetic tunnel junction.," *Nature,* vol. 613, pp. 490-495, 2023.

[7]  P. Qin, H. Yan, X. Wang, et al., "Room-temperature magnetoresistance in an all-antiferromagnetic tunnel junction.," *Nature,* vol. 613, pp. 485-489, 2023.

[8]  C. Boix-Constant, S. Jenkins, R. Rama-Eiroa, et al., "Multistep magnetization switching in orthogonally twisted ferromagnetic monolayers," *Nat. Mater.,* vol. 23, p. 212–218, 2024.

[9]  J. Nogués and I. K. Schuller, "Exchange bias.," *J. Magn. Magn. Mater.,* vol. 192, pp. 203-232, 1999.

[10] M. Ali, P. Adie, C. Marrows, et al., "Exchange bias using a spin glass.," *Nature. Mater.,* vol. 6, pp. 70-75, 2007.

[11] M. Testa-Anta, B. Rivas-Murias, and V. Salgueiriño, "Spin frustration drives exchange bias sign crossover in CoFe2O4-Cr2O3 nanocomposites.," *Adv. Funct. Mater.,* vol. 29, p. 1900030, 2019.

[12] J. Zhang, J. Yang, G. L. Causer, et al., "Realization of exchange bias control with manipulation of interfacial frustration in magnetic complex oxide heterostructures.," *Phys. Rev. B,* vol. 104, p. 174444, 2021.

[13] H. Ohldag, A. Scholl, F. Nolting, et al., "Correlation between exchange bias and pinned interfacial spins.," *Phys. Rev. Lett.,* vol. 91, p. 017203, 2003.

[14] H. K. Gweon, S. Y. Lee, H. Y. Kwon, et al., "Exchange bias in weakly interlayer-coupled van der Waals magnet Fe3GeTe2.," *Nano Lett.,* vol. 21, p. 1672–1678, 2021.

[15] D. Kim, S. Park, J. Lee, et al., "Antiferromagnetic coupling of van der Waals ferromagnetic Fe3GeTe2.," *Nanotechnology,* vol. 30, p. 245701, 2019.

[16] R. Zhu, W. Zhang, W. Shen, et al., "Exchange bias in van der Waals CrCl3/Fe3GeTe2 heterostructures.," *Nano Lett.,* vol. 20, p. 5030–5035, 2020.

[17] M. -H. Phan, V. Kalappattil, V. O. Jimenez, et al., "Exchange bias and interface-related effects in two-dimensional van der Waals magnetic heterostructures: Open questions and perspectives.," *J. Alloys Compd.,* vol. 937, p. 168375, 2023.

[18] G. Hu, Y. Zhu, J. Xiang, et al., "Antisymmetric magnetoresistance in a van der Waals antiferromagnetic/ferromagnetic layered MnPS3/Fe3GeTe2 stacking heterostructure.," *ACS Nano,* vol. 14, p. 12037–12044, 2020.

[19] T. M. J. Cham, R. J. Dorrian, X. S. Zhang, et al., "Exchange bias between van der Waals materials: tilted magnetic states and field-free spin–orbit-torque switching.," *Adv. Mater.,* p. 2305739, 2023.

[20] C. Wang, K. Zhu, Y. Ma, et al., "Magnetism control with enhanced hard magnetic temperature in heterostructures based on the van der Waals magnet.," *Phys. Rev. B,* vol. 108, p. 094408, 2023.

[21] J. Liang, S. Liang, T. Xie, et al., "Facile integration of giant exchange bias in Fe5GeTe2/oxide heterostructures by atomic layer deposition.," *Phys. Rev. Materials,* vol. 7, p. 014008, 2023.

[22] R. Zhao, Y. Wu, S. Yan, et al., "Magnetoresistance anomaly in Fe5GeTe2 homo-junctions induced by its intrinsic transition.," *Nano Res.,* vol. 16, p. 10443–10450, 2023.

[23] J. -Z. Fang, H. -N. Cui, S. Wang, et al., "Exchange bias in the van der Waals heterostructure MnBi2Te4/Cr2Ge2Te6.," *Phys. Rev. B,* vol. 107, p. L041107, 2023.

[24] Z. Ying, B. Chen, C. Li, et al., "Large exchange bias effect and coverage-dependent interfacial coupling in CrI3/MnBi2Te4 van der Waals heterostructures.," *Nano Lett.,* vol. 23, p. 765–771, 2023.

[25] T. Zhang, Y. Zhang, M. Huang, et al., "Tuning the exchange bias effect in 2D van der Waals ferro-/antiferromagnetic Fe3GeTe2/CrOCl heterostructures.," *Adv. Sci.,* vol. 9, p. 2105483, 2022.

[26] K. Gu, X. Zhang, X. Liu, et al., "Exchange Bias Modulated by Antiferromagnetic spin-flop transition in 2D van der Waals heterostructures.," *Adv. Sci.,* p. 2307034, 2024.

[27] H. Dai, H. Cheng, M. Cai, et al., "Enhancement of the coercive field and exchange bias effect in Fe3GeTe2/MnPX3 (X = S and Se) van der Waals heterostructures.," *ACS Appl. Mater. Interfaces,* vol. 13, p. 24314–24320, 2021.

[28] S. Albarakati, W. -Q. Xie, C. Tan, et al., "Electric control of exchange bias effect in FePS3–Fe5GeTe2 van der Waals heterostructures.," *Nano Lett.,* vol. 22, p. 6166–6172, 2022.

[29] L. Zhang, X. Huang, H. Dai, et al., "Proximity-coupling-induced significant enhancement of coercive field and Curie temperature in 2D van der Waals heterostructures.," *Adv. Mater.,* vol. 32, p. 2002032, 2020.

[30] M. Tang, J. Huang, F. Qin, et al., "Continuous manipulation of magnetic anisotropy in a van der Waals ferromagnet via electrical gating.," *Nat. Electron.,* vol. 6, p. 28–36, 2023.

[31] H. Zhang, R. Chen, K. Zhai, et al., "Itinerant ferromagnetism in van der Waals Fe5-xGeTe2 crystals above room temperature," *Phys. Rev. B,* vol. 102, p. 064417, 2020.

[32] A. R. Wildes, V. Simonet, E. Ressouche, et al., "Magnetic structure of the quasi-two-dimensional antiferromagnet NiPS3.," *Phys. Rev. B,* vol. 92, p. 224408, 2015.

[33] R. Basnet, A. Wegner, K. Pandey, et al., "Highly sensitive spin-flop transition in antiferromagnetic van der Waals material MPS3 (M = Ni and Mn).," *Phys. Rev. Materials,* vol. 5, p. 064413, 2021.

[34] J. Li, Y. Meng, J. -S. Park, et al., "Determination of the Fe magnetic anisotropies and the CoO frozen spins in epitaxial CoO/Fe/Ag(001)," *Phys. Rev. B,* vol. 84, p. 094447, 2011.

[35] W. Kim, E. Jin, J. Wu, et al., "Effect of NiO spin orientation on the magnetic anisotropy of the Fe film in epitaxially grown Fe/NiO/Ag(001) and Fe/NiO/MgO(001)," *Phys. Rev. B,* vol. 81, p. 174416, 2010.

[36] S. Mangin, D. Ravelosona, J. A. Katine, et al., "Current-induced magnetization reversal in nanopillars with perpendicular anisotropy.," *Nature Mater.,* vol. 5, p. 210–215, 2006.

[37] W.-G. Wang, M. Li, S. Hageman, et al., "Electric-field-assisted switching in magnetic tunnel junctions.," *Nature Mater.,* vol. 11, pp. 64-68, 2012.

[38] S. Ikeda, K. Miura, H. Yamamoto, et al., "A perpendicular-anisotropy CoFeB–MgO magnetic tunnel junction.," *Nature Mater.,* vol. 9, p. 721–724, 2010.

[39] K. Ning, H. Liu, L. Li, et al., "Tailoring perpendicular magnetic anisotropy with graphene oxide membranes.," *RSC Adv.,* vol. 7, pp. 52938-52944, 2017.

[40] R. Fujita, P. Bassirian, Z. Li, et al., "Layer-dependent magnetic domains in atomically thin Fe5GeTe2.," *ACS Nano,* vol. 16, p. 10545–10553, 2022.

[41] M. Schmitt, T. Denneulin, A. Kovács, et al., "Skyrmionic spin structures in layered Fe5GeTe2 up to room temperature.," *Commun. Phys.,* vol. 5, p. 254, 2022.

[42] C. Tan, W. Q. Xie, G. Zheng, et al., "Gate-controlled magnetic phase transition in a van der Waals magnet Fe5GeTe2.," *Nano Lett.,* vol. 21, p. 5599–5605, 2021.

[43] A. Scholl, F. Nolting, J. Stöhr, et al., "Exploring the microscopic origin of exchange bias with photoelectron emission microscopy.," *J. Appl. Phys.,* vol. 89, p. 7266–7268, 2001.

[44] R. H. Wade, "The determination of domain wall thickness in ferromagnetic films by electron microscopy.," *Proc. Phys. Soc.,* vol. 79, p. 1237, 1962.

[45] H. -H. Yang, N. Bansal, P. Rüßmann, et al., "Magnetic domain walls of the van der Waals material Fe3GeTe2," *2D Mater.,* vol. 9, p. 025022, 2022.

[46] U. Gradmann, "Magnetic surface anisotropies.," *J. Magn. Magn. Mater.,* Vols. 54-57, pp. 733-736, 1986.

[47] D. Givord, O. F. K. McGrath, C. Meyer, et al., "Interface magnetic anisotropy.," *J. Magn. Magn. Mater.,* Vols. 157-158, pp. 245-249, 1996.

[48] D. Wang, R. Wu and A. J. Freeman., "First-principles theory of surface magnetocrystalline anisotropy and the diatomic-pair model.," *Phys. Rev. B,* vol. 47, p. 14932, 1993.

[49] C. Liu, S. D. Bader, "Perpendicular surface magnetic anisotropy in ultrathin epitaxial Fe films.," *J. Vac. Sci. Technol. A,* vol. 8, p. 2727–2731, 1990.

[50] J. Okabayashi, J. W. Koo, H. Sukegawa, et al., "Perpendicular magnetic anisotropy at the interface between ultrathin Fe film and MgO studied by angular-dependent x-ray magnetic circular dichroism.," *Appl. Phys. Lett.,* vol. 105, p. 122408, 2014.

[51] D. Weller, Y. Wu, J. Stöhr, et al., "Orbital magnetic moments of Co in multilayers with perpendicular magnetic anisotropy.," *Phys. Rev. B,* vol. 49, p. 12888, 1994.

[52] N. Nakajima, T. Koide, T. Shidara, et al., "Perpendicular magnetic anisotropy caused by interfacial hybridization via enhanced orbital moment in Co/Pt multilayers: magnetic circular X-Ray dichroism study.," *Phys. Rev. Lett.,* vol. 81, p. 5229, 1998.

[53] S. Peng, M. Wang, H. Yang, et al., "Origin of interfacial perpendicular magnetic anisotropy in MgO/CoFe/metallic capping layer structures.," *Sci. Rep.,* vol. 5, p. 18173, 2015.

[54] B. -Y. Wang, J. Y. Ning, T. H. Li, et al., "Antiferromagnet-induced perpendicular magnetic anisotropy in ferromagnetic Co/Fe films with strong in-plane magnetic anisotropy.," *Phys. Rev. B,* vol. 105, p. 184415, 2022.

[55] E. J. Escorcia-Aparicio, H. -J. Choi, W. L. Ling, et al., "90° Magnetization Switching in Thin Fe Films Grown on Stepped Cr(001).," *Phys. Rev. Lett.,* vol. 81, p. 2144, 1998.

[56] A. Castellanos-Gomez, M. Buscema, R. Molenaar, et al., "Deterministic transfer of two-dimensional materials.," *2D Mater.,* vol. 1, p. 011002, 2014.

# Acknowledgements

This work is primarily supported by US Department of Energy, Office of Science, Office of Basic Energy Sciences, Materials Sciences and Engineering Division under Contract No. DE-AC02-05CH11231 (van der Waals heterostructures program, KCWF16). The operations of the Advanced Light Source at Lawrence Berkeley National Laboratory are supported by the Director, Office of Science, Office of Basic Energy Sciences, and U.S. Department of Energy under Contract No. DE-AC02–05CH11231. Q.L. and M.Y. acknowledge support from National Natural Science Foundation of China (Grants No. 12174364, 12104003, 12241406), the Fundamental Research Funds for the Central Universities (No. wk2310000104), the National Key Research and Development Program of China (No. 2023YFA1406400), the USTC Research Funds of the Double First-Class Initiative (No. YD2140002004) and Natural Science Foundation of Anhui Province (Grant No. 2308085Y04); X.Z. and Z.Q.Q. acknowledges King Abdullah University of Science and Technology (KAUST), Office of Sponsored Research (OSR) and under the Award No. ORA-CRG10-2021-4665; C.H. and Z.Q.Q. acknowledge support from Future Materials Discovery Program through the National Research Foundation of Korea (No. 2015M3D1A1070467) and Science Research Center Program through the National Research Foundation of Korea (No. 2015R1A5A1009962). X.H. acknowledges the support from SRC-JUMP ASCENT center. H.Z. thanks the support from the startup funding at Ningbo Institute of Materials Technology and Engineering, Chinese Academy of Science. R.R. acknowledges the Air Force Office of Scientific Research 2D Materials and Devices Research program through Clarkson Aerospace Corp under Grant No. FA9550-21-1-0460.

## Author contributions

T.Y.W. and Z.Q.Q. designed the experiments, analyzed data and wrote the paper. T.Y.W., Q.L. and M.Y. performed the XMLD measurements. T.Y.W. and Y.S. performed XMCD, electronic transport and atomic force microscopy measurements. T.Y.W., Y.S. and S.Y. performed PEEM measurements and sample preparations. X.C., X.H., H.Z. and R.R. provided support and advice for electronic transport and atomic force microscopy measurements. A.T.N., C.K. and P.C.S. provided support and advice for the XMCD and XMLD measurements at the ALS. A.S. provided support for the PEEM measurements at the ALS. M.F.C. provided support for sample preparations. All authors contributed to discussions and manuscript preparation.

## Data availability

The data that support the findings of this study are available from the corresponding author upon reasonable request.

## Competing interests

The authors declare no competing interests.

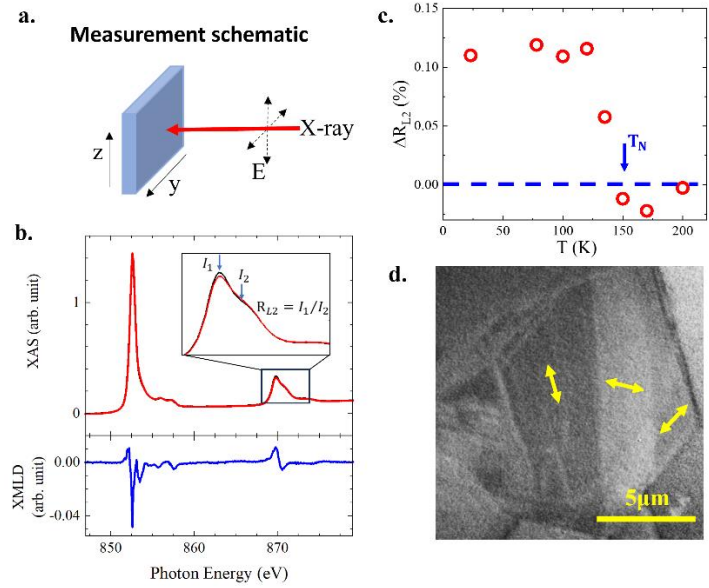

**Figure 1.** Direct characterization of antiferromagnetic orders in NiPS$_3$. **a:** Schematic of the XMLD characterization. Linearly polarized X-ray was incident onto sample surface at normal incidence. XAS at Ni L-edges were obtained via photoemission electron yield with X-ray polarizations along y and z directions, respectively, and XMLD was obtained from the difference of the two measurements. **b**: Ni XAS and XMLD at T = 23K. The $R_{L2}$ is defined as the ratio between X-ray absorption magnitudes $I_1$ and $I_2$ around Ni $L_2$ edge, and then the difference of $\Delta R_{L2}$ at the two X-ray polarizations is used for quantitative analysis of XMLD. **c:** Temperature-dependent $\Delta R_{L2}$. **d:** XMLD-PEEM image shows AFM domains in a ~70nm-thick exfoliated NiPS$_3$ flake. The three different contrasts (dark, grey, bright) correspond to the three equivalent Néel vector directions (labeled by the double-sided arrows) of vdW hexagonal monolayer lattice.

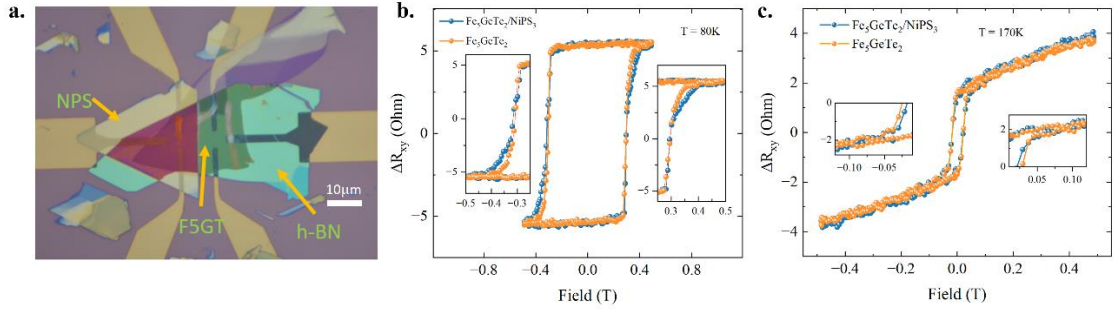

**Figure 2.** Temperature-dependent out-of-plane hysteresis loops via AHE of the same $Fe_5GeTe_2$(16nm) flake with and without $NiPS_3$. **a:** Sample image under optical microscope. 16nm-thick $Fe_5GeTe_2$ was transferred onto the Hall bars with 6 electrodes. The left half of the $Fe_5GeTe_2$ is covered by $NiPS_3$ so that one pair of Hall bar measures the $Fe_5GeTe_2/NiPS_3$ heterostructure and the other pair of Hall bar measures the single $Fe_5GeTe_2$ region. An h-BN flake serves as a capping layer on top of the whole device. **b-c:** Hall resistances at 80K and 170K by the two hall bars with magnetic field sweeping along out-of-plane direction after 9T out-of-plane field cooling. Zoom-in insets focus on the switching behavior near the coercivities. At 80K, the hysteresis loop of the hetereostructure region shows the two-step switching process while the 80K loop of the single flake region and the 170K loops of both regions do not.

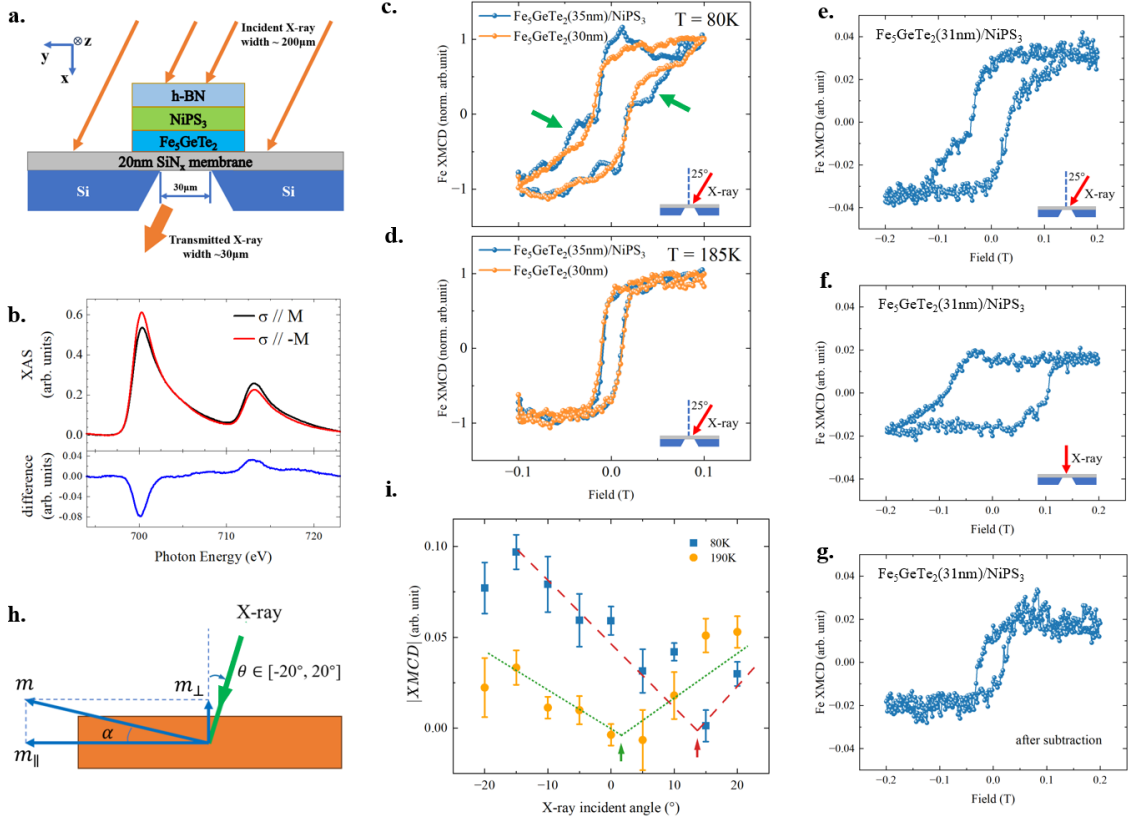

**Figure 3.** Temperature-dependent XMCD hysteresis loops and angle-dependent XMCD amplitudes measured on micron-sized $Fe_5GeTe_2$(30~35nm)/$NiPS_3$ heterostructure using X-ray with way larger beam size, via the membrane window method (see Methods section). **a:** Schematic of the measurement geometry of the membrane window method. The 30μm window serves as an aperture to greatly enhance the signal-to-noise ratio of the macroscopic X-ray techniques on micron-sized vdW samples. **b:** $Fe_5GeTe_2$ XAS and XMCD measured at Fe L-edges. **c-d**: XMCD hysteresis loops measured with 25° off-normal incident X-rays at Fe $L_3$ edge on $Fe_5GeTe_2$(35nm)/$NiPS_3$(31nm) heterostructure and $Fe_5GeTe_2$(30nm) single flake, respectively. The 25° loops exhibit easy-axis shape similar to that of the 51nm sample from Supplementary Figure S2b with extra-step switching process. Arbitrary units of the vertical axes are normalized for comparison between the loop shapes of the two different samples. **e-f:** XMCD hysteresis loops measured on another $Fe_5GeTe_2$(31nm)/$NiPS_3$ heterostructure at both 25° and 0° incident angles, respectively, to better reveal both the out-of-plane and in-plane components; **g**: Calculated in-plane loop obtained by subtracting **f** from **e** with $cos(25°)$ projecting factor. **h:** Schematic of the angular dependent XAS measurement for quantitatively determining the in-plane and out-of-plane components of $Fe_5GeTe_2$ in the heterostructure. $m_\perp$ and $m_\parallel$ denote the $Fe_5GeTe_2$ magnetic moment of the interfacial region and the region away from the interface, respectively; $m$ denotes the total magnetic moment with angle $\alpha$ from the sample plane; The signed incident angle $\theta$ of the X-ray ranges from $-20°$ to $20°$. The sample was cooled in a -1.8T field along $\theta = -45°$. After the field is turned off, the magnetic moments are supposed to return to their own easy axes. **i:** Angular dependent $|XMCD|$ at 80K and 190K after the 1.8T field cooling. The red dashed line and the green dotted line are guide to the eyes for the relationship $|sin(\theta - \alpha)|$ for small angles. The two upward arrows show approximately at which angle the XMCD magnitude reaches zero for the two temperatures.

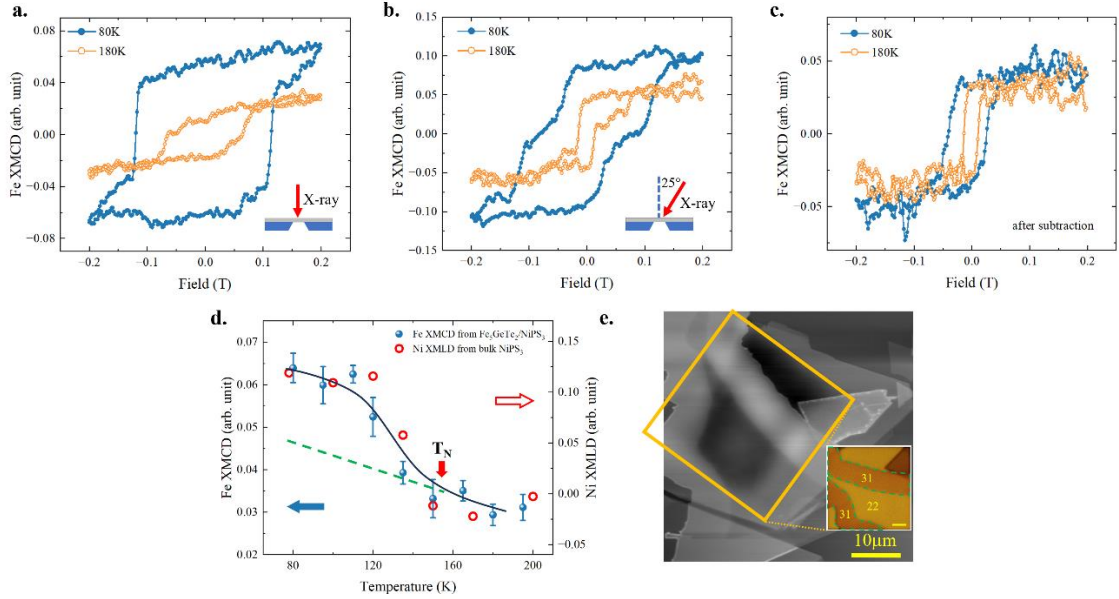

**Figure. 4.** Temperature-dependent XMCD hysteresis loops measured on $Fe_5GeTe_2$(one flake with two regions with 22nm and 31nm thicknesses)/$NiPS_3$ heterostructure on membrane window at **a:** 0° and **b**: 25° incident angle of the X-rays, respectively. **c:** Calculated in-plane hysteresis loop obtained by subtracting **a** from **b** considering the cos (25°) projection factor. **d:** Blue dots with error bar depict the temperature-dependent "saturation" (Fe XMCD amplitude near 0.2T) of the out-of-plane loops. Red open circles are the temperature-dependent Ni XMLD data from Fig. 1c serving as a reference to compare with the temperature-dependence of out-of-plane Fe XMCD. The solid curve through the data points is a guide to the eye, and the straight dashed line is the extrapolation of the high-temperature Fe XMCD data to the low-temperature region for quantitative analysis. Generally, the temperature-dependence of the magnetization of a ferromagnet below the Curie temperature can be anything far more complicated than linear relationship, so one cannot use only the temperature-dependence for a rigorous linear extrapolation. However, since the temperature range of the measurement (80K-190K) is far below the Curie temperature of the $Fe_5GeTe_2$ (~280K), we used the linear extrapolation from the linear fitting of 5 points with the highest temperatures as a guide to the eye to better illustrate our analysis based on the spin-twisting state hypothesis. The fitted slope and intercept are (-13 ± 5)×$10^{-5}$ and 0.056 ± 0.009, respectively. **e:** Atomic force microscopy image of the sample on the membrane window substrate. The yellow square shows the location of the membrane window. The inset labels the 31nm and 22nm regions of the $Fe_5GeTe_2$ flake, with the scale bar corresponding to 5μm. The area ratio between the 31nm region and the 22nm region was calculated to be 1.44 : 1.

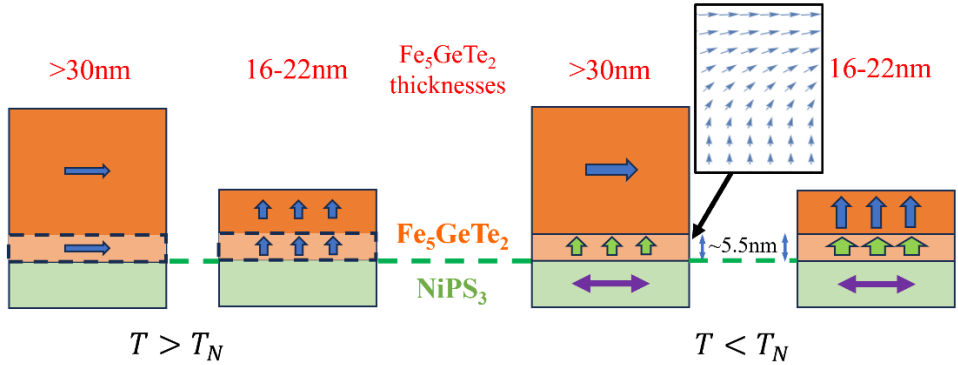

**Figure 5.** Schematic of magnetizations of the Fe$_5$GeTe$_2$/NiPS$_3$ heterostructures in this work. At high temperature without the interfacial spin frustration, the Fe$_5$GeTe$_2$ magnetization exhibits its intrinsic property of having an in-plane magnetization at thicker thickness and out-of-plane magnetization at thinner thickness. Below the NiPS$_3$ Néel temperature, the interfacial spin frustration between Fe$_5$GeTe$_2$ and the C-type AFM NiPS$_3$ induces a interfacial perpendicular magnetization (green arrow) of the Fe$_5$GeTe$_2$ of ~5.5nm thickness which switches separately from the bulk of the Fe$_5$GeTe$_2$ magnetization (blue arrow) away from the interface, leading to a two-step switching process as shown in the hysteresis loops as well as the unconventional twisting spin structure along the out-of-plane direction.

# Supplementary Materials

**About Ni XMLD in bulk NiPS$_3$ crystal and Fe$_5$GeTe$_2$(35nm)/NiPS$_3$ heterostructure**

It should be mentioned that the XMLD result of bulk NiPS$_3$ in main text Fig. 1 does not depend on whether a 0.4T magnetic field (the maximum magnetic field available for the sample holder used at BL4.0.2) was applied or not during the cooling process, which is expected because 0.4T field is way smaller than the spin-flop transition field (~6T) of NiPS$_3$ [1]. However, it is surprising to observe a non-zero macroscopic XMLD in bulk NiPS$_3$ because the existence of AFM domains should usually average out the macroscopic XMLD signal. We notice that NiPS$_3$ bulk crystals are much softer and more prone to form wrinkles compared to graphite (HOPG) and other vdW crystals during mechanical exfoliation. Therefore, we believe that the pre-cleaving of our bulk NiPS$_3$ crystal with tapes had induced a large-scale uniaxial strain over the crystal surface which favors a particular AFM spin axis among the three equivalent spin axes, leading to a non-zero macroscopic XMLD signal in our bulk sample, similar to the strain-induced AFM domains in cleaved NiO single crystal [2].

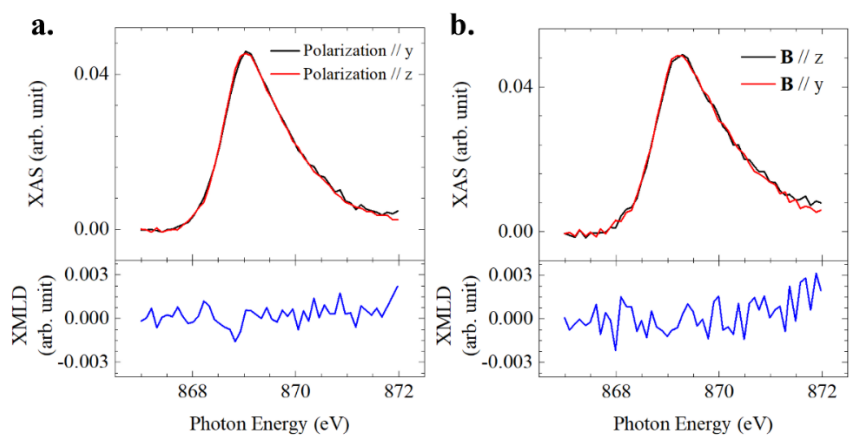

**Figure S1.** Ni XAS and XMLD of the Fe$_5$GeTe$_2$(35nm)/NiPS$_3$ flake measured by the membrane window substrate method. (See Methods section.) Linear polarized X-ray was incident onto sample surface with normal incidence. Ni L-edge XAS was obtained via transmitted X-ray, with either two perpendicular field directions or two perpendicular X-ray polarization directions, and XMLD was obtained from the difference of each two measurements: **a**: The 0.4T magnetic field was fixed along the y-axis and the X-ray linear polarization was switched between y- and z- axes. **b:** The X-ray linear polarization was fixed along z-axis and the 0.4T magnetic field was switched between y- and z- axes. The NiPS$_3$ in the heterostructure does not show an overall XMLD effect within error bar. This shows that the NiPS$_3$ AFM order was not aligned by the 0.4T field and the interfacial coupling with Fe$_5$GeTe$_2$.

We characterized the NiPS$_3$ XMLD in the Fe$_5$GeTe$_2$(35nm)/NiPS$_3$(31nm) heterostructure. The 35nm-thick Fe$_5$GeTe$_2$ film was chosen because it exhibits an intrinsic in-plane magnetization as is shown by our XMCD measurement in the main text. The interfacial coupling between in-plane FM Fe$_5$GeTe$_2$ and in-plane AFM NiPS$_3$ should best explore the effect of interfacial spin frustration. After cooling the Fe$_5$GeTe$_2$(35nm)/NiPS$_3$(31nm) sample from room temperature to 80K within a 0.4T in-plane magnetic field, the Ni XMLD was measured at normal X-ray incidence. Contrary to the bulk NiPS$_3$, the micron-sized Fe$_5$GeTe$_2$/NiPS$_3$ thin film heterostructure shows no measurable Ni XMLD signal at 80K (Fig. S1a). Noting that the NiPS$_3$ layer in Fe$_5$GeTe$_2$(35nm)/NiPS$_3$(31nm) sample was exfoliated from the same bulk NiPS$_3$ crystal in Fig. 1, the absence of XMLD signal in Fe$_5$GeTe$_2$(35nm)/NiPS$_3$(31nm) indicates that the NiPS$_3$ thin layer should consist of multiple AFM domains as shown in Fig. 1d even after the 0.4T field cooling. We further measured the Ni XMLD at 80K by switching the Fe$_5$GeTe$_2$ magnetization along two orthogonal in-plane directions (Fig. S1b). Again, no measurable XMLD was observed, showing that the Fe$_5$GeTe$_2$/NiPS$_3$ interfacial magnetic coupling is not strong enough to align/rotate the NiPS$_3$ Néel vector either. Our result shows that unlike the NiPS$_3$ bulk crystal where a uniaxial strain could be induced over a macroscopic scale (greater than ~100μm) by surface cleaving, the micron-sized exfoliated and PDMS-transferred NiPS$_3$ thin flakes have relaxed any macroscopic uniaxial strain to result in multiple AFM domains with their local spin axes following the local magnetocrystalline anisotropy axes. The local magnetic anisotropy must be strong enough against spin-flop transition under the 0.4T cooling field as well as against the Fe$_5$GeTe$_2$/NiPS$_3$ interfacial magnetic coupling.

# XMCD Hysteresis loops of a 51nm-thick Fe$_5$GeTe$_2$ single flake

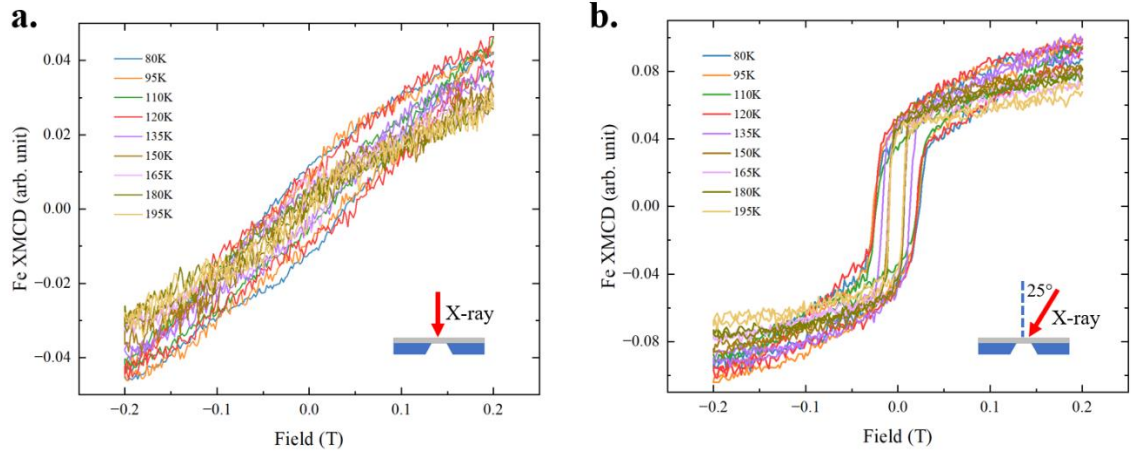

**Figure S2**. XMCD Hysteresis loops of Fe$_5$GeTe$_2$(51nm) single flake on membrane window substrate. **a:** Out-of-plane hysteresis loops measured with normally incident X-rays. **b:** Hysteresis loops measured with 25° off-normal incident X-rays. The characteristic hard-axis out-of-plane hysteresis loop and the easy-axis in-plane hysteresis loop show that the 51nm-thick Fe$_5$GeTe$_2$ has an in-plane easy axis from 80K to 195K.

**In-plane and out-of-plane field cooling processes for $Fe_5GeTe_2$(one flake with 22nm and 31nm thicknesses)/$NiPS_3$**

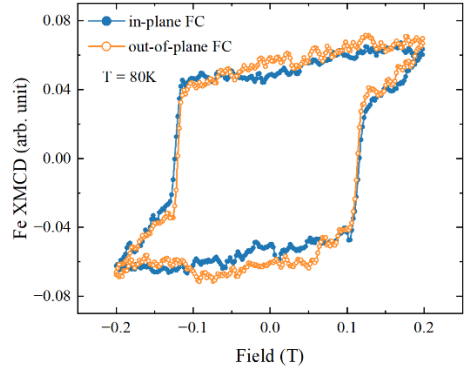

**Figure S3.** 80K out-of-plane hysteresis loops of the $Fe_5GeTe_2$(one flake with two regions, 22nm and 31nm thicknesses)/$NiPS_3$ heterostructure on membrane window after 1.8T in-plane field cooling and 1.8T out-of-plane field cooling, respectively.

An interesting question is whether the out-of-plane 22nm $Fe_5GeTe_2$ will show any difference after an out-of-plane field cooling process. We cooled down the sample from 200K down to 80K with an out-of-plane magnetic field of 1.8T, and then measured the out-of-plane hysteresis loop at 80K. As shown in Fig. S3, the obtained hysteresis loop after out-of-plane field cooling is identical to that after in-plane field cooling in Fig. 4a, showing that out-of-plane 1.8T field cooling process does not change the interfacial coupling effects on the out-of-plane hysteresis loops (i.e. absence of the training effect), further supporting our previous analysis that the interfacial coupling does not affect the $NiPS_3$ orders, and that the $NiPS_3$ samples only contain fully compensated spins at the interface which is expected from a C-type AFM.

**Spin-flop-like effects of the interfacial spin frustration in Fe$_5$GeTe$_2$/NiPS$_3$**

Consider a simple model of FM/AFM bilayers with the films parallel to the xy-plane (the interface normal direction is along the z-axis), the FM exchange interaction of $-J_0 \sum \vec{S}_{F,i} \cdot \vec{S}_{F,j}$ ($J_0$ is the FM-FM nearest-neighbor interaction), and the interfacial AFM-FM interaction of $-J_1 \sum \vec{S}_{AF,i} \cdot \vec{S}_{F,i}$ ($J_1$ is the AFM-FM interfacial interaction). Without losing generality, we assume that the AF spin spins are parallel to the y-axis in the film plane and alternates their directions between +y and -y for every lattice displacement of $\Delta x = a/2$. The AFM-FM interfacial interaction obviously favors the local FM spin parallel to the local AFM spin along the $\pm y$ directions while the FM-FM interaction favors parallel alignment of all FM spins in the same direction, leading to the spin frustration in the FM layer. Taking the fact that the FM magnetization is almost uniform ($J_1 \ll J_0$), we consider the FM energy as a function of its orientation in the yz-plane ($\theta_i$ being the local polar angle of the FM spin away from the z-axis in the yz plane). For perfectly uniform FM magnetization ($\theta_i = \theta = const.$), the interfacial FM-AFM interaction leads to a $\theta$-independent interfacial energy because the interfacial interaction of the FM spin with one AFM sublattice is exactly balanced by the interfacial interaction of FM spin with another AFM sublattice, $E = -J_1 \cos\left(\frac{\pi}{2} - \theta\right) - J_1 \cos\left(\frac{\pi}{2} + \theta\right) = 0$. Allowing a small twisting of the FM local spin direction towards the local AFM spin direction from its average direction ($\theta_i = \bar{\theta} \pm \delta\theta$, $\delta\theta \ll 1$, where the $\pm$ corresponds to $\vec{S}_{AF,i} = \pm \hat{y}$, respectively), the FM-AFM interfacial interaction energy can be lowered by $-J_1 \cos\left(\frac{\pi}{2} - \bar{\theta} - \delta\theta\right) - J_1 \cos\left(\frac{\pi}{2} + \bar{\theta} - \delta\theta\right) = -J_1 \sin(\bar{\theta} + \delta\theta) + J_1 \sin(\bar{\theta} - \delta\theta)$. Such local FM spin twisting, however, will increase the FM-FM interaction energy by $J_0 - J_0 \cos(2\delta\theta) = 2J_0 \sin^2(\delta\theta)$. These two competing terms (spin frustration) lead to a total energy change per site of

$$E = -\frac{J_1[\sin(\bar{\theta} + \delta\theta) - \sin(\bar{\theta} - \delta\theta)]}{2} + J_0 \sin^2(\delta\theta)$$

$$\approx -J_1 \cos\bar{\theta}\, \delta\theta + J_0 (\delta\theta)^2 \qquad (1)$$

Minimizing the energy leads to

$$\delta\theta \approx \sin(\delta\theta) = \frac{J_1}{2J_0} \cos\bar{\theta} \qquad (2)$$

$$E_{min} = -\frac{J_1^2}{4J_0} \cos^2\bar{\theta} \qquad (3)$$

Eqn. (3) corresponds to a uniaxial magnetic anisotropy with the easy axis perpendicular to the film ($\bar{\theta} = 0$). From the above discussion, it is clear that the interfacial spin frustration results in a spin-flop-like effect with a small angular twisting of the FM spins relative to the AFM spins.

It is constructive to compare the result of Eqn. (3) to the well-known spin-flop transition in AFM to better understand the physical meaning of frustration-induced magnetic anisotropy. In the spin-flop transition, applying a magnetic field along the easy axis of an AFM will switch the AFM spin axis by 90° to be perpendicular to the magnetic field. This transition occurs by a small canting (or twisting) of the AFM spins, which creates a small amount of net spin perpendicular to the AFM spin axis, that the Zeeman interaction between this net spin and the magnetic field lowers the total energy by aligning the net spin to the magnetic field direction (e.g., perpendicularly alignment of the AFM spin axis to the magnetic field direction). From the point of view of mean field theory, the FM-AFM interaction $-J_1 \sum \vec{S}_{AF,i} \cdot \vec{S}_{F,i} \approx -J_1 \sum \vec{S}_{AF,i} \cdot \langle \vec{S}_{F,i} \rangle$ is equivalent to a "magnetic field" of $\vec{H} = J_1 \langle \vec{S}_{F,i} \rangle$ applied to the AFM spins at the interface $-J_1 \sum \vec{S}_{AF,i} \cdot \langle \vec{S}_{F,i} \rangle \approx -\sum \vec{S}_{AF,i} \cdot \vec{H}$. Therefore, the final state should have the AFM spin axis perpendicular to the "magnetic field" direction which is the FM spin direction, except that in this case the AFM spins are fixed in space and the "magnetic field" $\vec{H} = J_1 \langle \vec{S}_{F,i} \rangle$ can rotate freely in space. Then, the spin-flop transition in AFM is equivalent to the rotation of the "magnetic field" $J_1 \langle \vec{S}_{F,i} \rangle$ to the perpendicular direction of the AFM spins in the FM/AFM system plus a small twisting of the "magnetic field" (FM magnetization) relative to the AFM spins.

**Comparison of spin frustration effect between FM/AFM heterostructures and spin liquid/glass systems**

Spin frustration has been traditionally studied in spin glass and spin liquid systems in which spin frustration is "intrinsic" in the sense that it exists in every group of three neighboring spins, resulting in multiple energy-degenerate states in each neighboring group as well as the absence of long-range order of spin alignment. Consequently, low-energy quasiparticle single-spin excitations could exist globally known as the quantum spin liquid state. Therefore, it is natural for these systems to have notable slow spin relaxation processes due to the highly non-convex energy functions near the "ground state points" in the state space.

On the contrary, FM system exhibits long-range order of spin alignments due to consistently uniform intrinsic ferromagnetic exchange coupling, and thus the spin frustration in FM/AFM systems (e.g. $Fe_5GeTe_2$/$NiPS_3$) originates from the interfacial coupling through spin-flop-like interactions and is purely extrinsic. The direct consequence of such extrinsic spin frustration is to induce a uniform phenomenological perpendicular magnetic anisotropy (PMA) over the whole FM surface. Such FM systems with extrinsic PMA only exhibit two well-defined long-range ordered magnetic ground states (either spin-up or spin-down), meaning that the energy function is well convex near the ground state points in the state space. Therefore, the key factor in spin frustration in FM/AFM systems is to understand the origin of the frustration-induced global magnetic anisotropy rather than a notable slow spin relaxation progress. This is drastically different from the spin frustration effects in spin liquid or spin glass systems.

**Reference**

[1] R. Basnet, A. Wegner, K. Pandey, et al., "Highly sensitive spin-flop transition in antiferromagnetic van der Waals material MPS3 (M = Ni and Mn).," *Phys. Rev. Materials,* vol. 5, p. 064413, 2021.

[2] S. Mandal, K. S. R. Menon, F. Maccherozzi, et al., "Strain-induced nonequilibrium magnetoelastic domain structure and spin reorientation of NiO(100)," *Phys. Rev. B,* vol. 80, p. 184408, 2009.